\documentclass{sig-alternate}
\usepackage{algorithm2e}
\usepackage{graphicx}
\usepackage{rotating}
\usepackage{booktabs}
\usepackage{color}
\usepackage[table]{colortbl}

\def\Stryker{{\textsf{Stryker}}}
\def\TACO{{\textsf{TACO}}}
\def\muJava{{\textsf{muJava}}}

\begin{document}


\title{Stryker: Scaling Specification-Based Program Repair by Pruning Infeasible Mutants with SAT}

\numberofauthors{3}

\author{
\alignauthor Luciano Zem\'{\i}n\\
	\affaddr{Buenos Aires Institute of Technology (ITBA), Argentina}
\alignauthor Sim\'on Guti\'errez Brida\\
	\affaddr{University of Rio Cuarto and CONICET, Argentina}
\alignauthor Santiago Berm\'udez\\
	\affaddr{Buenos Aires Institute of Technology (ITBA), Argentina}
    \and
\alignauthor Santiago~Perez~De~Rosso\\
	\affaddr{MIT CSAIL, Cambridge, MA, USA}
\alignauthor  Nazareno Aguirre\\
	\affaddr{University of Rio Cuarto and CONICET, Argentina}
\alignauthor  Ali Mili\\
	\affaddr{Dept.~of Computer Science, NJIT, Newark, NJ, USA}
    \and
\alignauthor Ali Jaoua\\
	\affaddr{Dept.~of Computer Science, Qatar University, Qatar}
\alignauthor Marcelo F.~Frias\\
	\affaddr{Buenos Aires Institute of Technology (ITBA) and CONICET, Argentina}
}

\maketitle

\begin{abstract}
Many techniques for automated program repair involve syntactic program transformations. Applying combinations of such transformations on faulty code yields fix candidates whose correctness must be determined. Exploring these combinations leads to an explosion on the number of generated fix candidates that severely limits the applicability of such fault repair techniques. This explosion is most times tamed by not considering fix candidates exhaustively, and by disabling intra-statement modifications. In this article we present a technique for program repair that considers an ample set of intra-statement syntactic operations, and explores fix candidates exhaustively up to a provided bound. The suitability of the technique, implemented in our tool \Stryker{}, is supported by a novel mechanism to detect and prune infeasible fix candidates. This allows \Stryker{} to repair programs with several bugs, whose fixes require multiple modifications. We evaluate our technique on a benchmark of faulty Java container classes, which \Stryker{} is able to repair, pruning significant parts of the space of generated candidates when more than one bug is present in the code.
\end{abstract}

\section{Introduction}


The significant advances in automated analysis techniques have led, in the last few decades, to the development of powerful tools able to assist software engineers in software development, that have proved to greatly contribute to software quality. Tools based on model checking \cite{Clarke+2000}, constraint solving \cite{Nieuwenhuis+2006}, evolutionary computation \cite{DeJong2006} and other automated approaches, are being (successfully) applied to various aspects of software development, from requirements specification \cite{BharadwajHeitmeyer1999,Fuxman+2001} to verification \cite{JhalaMajumdar2009} and bug finding \cite{JacksonVaziri2000,Galeotti+2013}. The emphasis of these tools and techniques has been, however, to detect the existence of defects in software artifacts and specifications, via various approaches, including bug finding based on testing \cite{Gligoric+2010}, runtime analysis \cite{Chen+2008} and static analysis \cite{Bessey+2010}. Since the seminal work of Arcuri and Yao \cite{ArcuriYao2008}, some of these techniques have started to be applied to \emph{repair} software, through the automation of activities such as fault localization \cite{JonesHarrold2005} and fault correction \cite{DW10,KNSK13,WNLF09}.

Even though the idea of automated bug fixing is appealing, automatically fixing arbitrary program defects is known to be infeasible. Therefore, automated program repair must necessarily sacrifice completeness. Several effective techniques for program repair resort to exploring a large (but limited) set of fix candidates obtained via syntactic modifications to a faulty program. Moreover, for these techniques to scale reasonably, the space of fix candidates must often be tamed, by limiting the set of syntactic modifications considered (e.g., no intra-statement modifications), or not exhaustively exploring all (bounded) candidates (e.g., using a genetic algorithm instead of exhaustive search).  For example, {\em GenProg} \cite{LeGoues+2012} uses evolutionary computation to syntactically evolve a program until an acceptable fix is found. Each candidate repair (syntactic modification) is applied to the original program to produce a new program whose fitness is evaluated using a test suite. Intra-statement syntactic modifications are not considered so as to limit the candidate space, and the fitness function is used to maintain a reduced population of candidates throughout the evolutionary computation process.  In \emph{PAR} \cite{KNSK13}, only certain program modifications are considered, which are learned from human-written patterns. Thus, the number of candidates to be considered as fixes is significantly reduced, which also reduces the kind of errors the approach is able to fix.

Intra-statement program modifications, i.e., those that alter the expressions within statements, are generally not considered by effective approaches to program repair. A main limitation in considering such program modifications, or as we call them, mutations, is the explosion of fix candidates, as previously described. 
        Thus, approaches that take into account such program modifications, limit these to a very reduced set of mutations (e.g., \cite{GMK11}), reducing the class of bugs the corresponding techniques are able to deal with.

In this paper, we present a technique for program repair that enables us to overcome this limitation. The technique, implemented in our tool \Stryker{}, considers an ample set of intra-statement syntactic operations, and explores fix candidates exhaustively up to a provided bound. It uses a mutation generation tool to produce fix candidates, and combines runtime analysis and bounded verification to assess the candidates' suitability. Moreover, the technique introduces a mechanism to detect and prune infeasible fix candidates, that allows it to repair programs with errors whose fixes require multiple modifications, and even to repair programs with several errors. We 
perform a careful assessment of the effectiveness and limitations of the technique, and provide an empirical analysis of its effectiveness, based on a benchmark of increasingly complex collection implementations.  


The remainder of the article is organized as follows. In Section~\ref{program-repair} we discuss program repair based on syntactic modifications, a family of techniques comprising our approach. In Section~\ref{ourApproach} we describe the scope of our technique, and its main characteristics. In Section~\ref{stryker} we describe its realization in the \Stryker{} tool. In Section \ref{pruningSection} we describe the pruning mechanism, a main contribution of this article. In Section~\ref{related} we discuss related work. In Section~\ref{eval} we evaluate the technique experimentally and, finally, in Section~\ref{conclu}, we present our conclusions and proposals for further work.

\section{Program Repair based on Syntactic Modification Operations}
\label{program-repair}

Program repair based on syntactic operations for program modification is a family of techniques for automatically repairing faulty programs through the application of transformations that modify the program's code. In its general form, the technique can be described as an exhaustive search that, given a program specification, a faulty program to repair and a fixed set of syntactic operators, that we also call mutation operations: \emph{(i)} takes the faulty program to be repaired as the initial repair candidate; \emph{(ii)} if $p$ is a repair candidate, and $q$ is the result of applying a mutation operator on $p$, then $q$ is also a repair candidate; and \emph{(iii)} a candidate $s$ is \emph{successful} if it satisfies the provided specification.

This problem statement makes it clear that the space of repair candidates depends on the number $b$ of mutation operators to be considered (the branching factor), and the maximum number $d$ of successive mutations considered to generate the candidates (the depth of the solution). A geometric sum explains a search space consisting of $\frac{b^{d+1}-1}{b-1}$ ($O(b^d)$) candidates. Consider for instance method \texttt{getNode}, shown in Alg.~\ref{sample}. 
Table~\ref{mutExplosion} shows that even for such a small piece of code and a modest number of operators, the number of mutants generated can grow to an extent that makes visiting it infeasible. Existing tools tame this explosion in different ways. Some reduce the explosion by bringing down the branching factor, using a single mutation (e.g., \cite{GMK11}), considering a very small set of mutators (e.g., based on patterns of human-written fixes \cite{KNSK13}), or considering coarse grained mutations (e.g., no intra-statement program modifications \cite{LeGoues+2012}). Others resort to non exhaustive heuristic search, e.g., based on evolutionary computation \cite{LeGoues+2012}. We will further discuss these approaches in Section~\ref{related}. 
\SetKwBlock{Body}{}{end}
\LinesNumbered
\begin{algorithm}[t]
\small
$\mathbf{SListNode}\ \mathrm{getNode}(\mathbf{int}\ \mathit{i})$ 
	\Body{
		$\mathbf{SListNode}\ \mathit{current} = \mathit{this.head}$\;
		$\mathbf{SListNode}\ \mathit{result} = \mathsf{null}$\;
		$\mathbf{int}\ \mathit{current\_index} = 0$\;
		\While{$(\mathit{result} == \mathsf{null}\ \&\&\ \mathit{current}\mbox{ \emph{!=} }\mathsf{null})$}{
			\If{$(\mathit{i} == \mathit{current\_index})$}{
				$\mathit{result} = \mathit{current}$\;
			}
			$\mathit{current\_index} = \mathit{current\_index} + 1$\;
			$\mathit{current} = \mathit{current.next}$\;
		}
		$\mathbf{return}\ \mathit{result}$\;  
	}
\normalsize
\caption{A sample method: \texttt{getNode}.}\label{sample}
\end{algorithm}

\begin{table}[t]
\begin{center}
\small
\begin{tabular}{c r}
Search Depth                            &	No. of Mutants (Fix Candidates)        \\
\hline
1 					&	40		                               	\\
2 					&	1,604			                \\
3 					&	64,684		        	        \\
4 					&	$>$ 20 million		                
\end{tabular}
\normalsize
\end{center}
\caption{Mutants generated from a \texttt{getNode} method mutating 4 faulty lines using 18 mutators, as search depth increases.}
\label{mutExplosion}
\end{table}

The above general problem statement requires some mechanism to establish whether a program satisfies its specification or not. Some approaches to program repair use tests as specifications (e.g., \cite{KNSK13,LeGoues+2012}), considering a program correct if it passes all tests in a provided suite. 
Other approaches, e.g. \cite{GMK11}, require a logical specification of the program to repair given in terms of a pre- and a post-condition, and employ a verification tool to decide whether a candidate is an acceptable fix or not. In particular, bounded verification tools can be used to make such checking fully automated.

\section{Our Program Repair Approach}\label{ourApproach}


As correctly pointed out in \cite{M14}, the class of defects a particular program repair approach tackles needs to be explicitly stated. Since 
we deal with automated program repair, let us then precisely formulate the scope of the program repair problem our technique enables us to tackle, including the class of bugs that we aim to automatically repair. 

\medskip
\textbf{Goal:} \emph{To provide an automated and efficient technique for repairing programs annotated with specifications (given in terms of pre- and post-conditions) by correcting errors resulting as a consequence of the simultaneous occurrence of a 
number of syntactic mistakes within program statements.}

\medskip

Let us remark that we require programs to be equipped with specifications, given in terms of pre- and post-conditions. Thus, a program will be considered incorrect 
if it does not satisfy its specification. Also, the considered syntactic mistakes, or \emph{mutations} (following \cite{Ma+2005}), appear \emph{within} statements. In particular, we do not aim at repairing source code whose faults are due to statements that are missing 
in the code (although some of these can be overcome through our intra-statement program repair). 

\subsection{Tasks Involved in Our Approach}

\begin{figure}[t]
\begin{center}
\includegraphics[width=1.0\columnwidth]{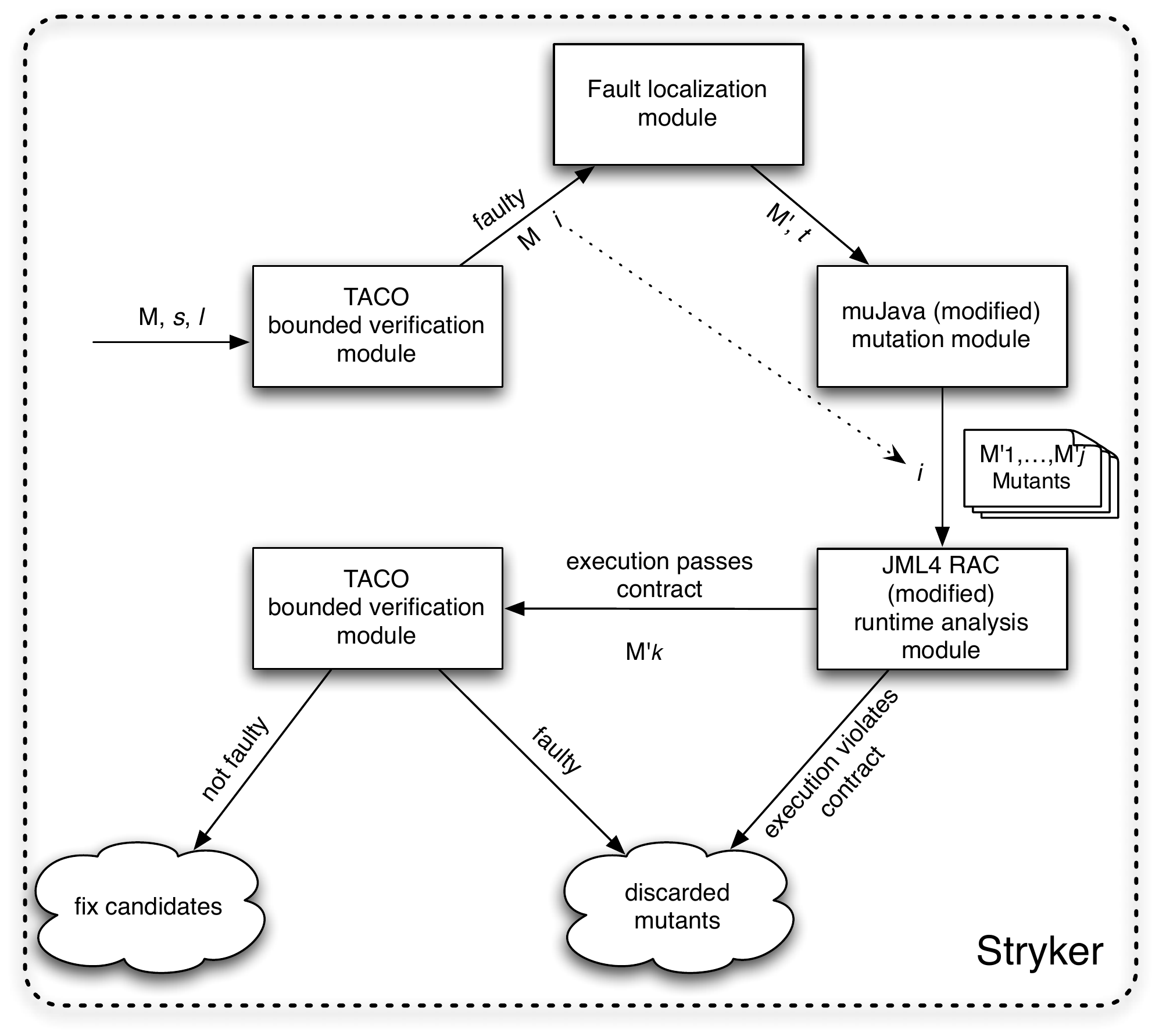}
\end{center}
\caption{A schematic description of \Stryker.}
\label{architecture}
\end{figure}

Our approach to program repair is depicted in Figure~\ref{architecture}. We describe here the main technologies involved in the technique's components. We will further describe how these components are actually realized in the following section. 

\textbf{Fault Detection.} Our technique applies to contract-equipped source code. To achieve full automation in fault detection, 
we consider \emph{bounded correctness}: a program will be viewed as incorrect if, given a user provided \emph{scope} (establishing a maximum number of objects, ranges for numerical types and maximum number of iterations in the program), there exists a program state within this scope satisfying the precondition, such that if the program under analysis is executed in it, it terminates within the iteration scope, in a state that does not satisfy the post-condition. Bounded correctness is decided automatically using \TACO{} \cite{Galeotti+2013}, which reduces bounded verification to boolean satisfiability, and employs off-the-shelf SAT solvers.  


\textbf{Fault Localization.}
Once a program is determined to be faulty, a mechanism for identifying the lines of code that are likely to be blamed for the fault is necessary. Our technique does not depend on any particular fault localization approach, and any one may be used for this matter. The fault localization module used in this paper is based on \emph{unsatisfiable cores}~\cite{LynceMarquesSilva2004}. Given an unsatisfiable propositional CNF formula $\alpha$, an unsat core $\textit{core}_{\alpha}$ of $\alpha$ is a subformula of it (subset of its set of clauses) that is also unsatisfiable. Intuitively, it can be thought of as the \emph{cause} of the unsatisfiability of $\alpha$, since $\alpha$ is unsatisfiable \emph{because} it is a conjunction involving an unsatisfiable conjunct $\textit{core}_{\alpha}$. Some modern SAT solvers have the ability of computing unsat cores from unsatisfiability proofs (i.e., runs of the SAT solver that led to unsatisfiability). 



\textbf{Fix Candidates Generation.}
Once the suspicious lines are identified, syntactical variants of the faulty program are produced using an extension of the mutation testing tool MuJava \cite{Ma+2005}. 
Our approach then differs substantially from the approaches mentioned in the previous section in the way it produces fix candidates. 
By using (an extension of) MuJava, we consider an ample set of intra-statement mutation operations, and we disregard the insertion and deletion of program statements. 
Intra-statement program alterations are the mutations employed in mutation testing \cite{PezzeYoung2008}, 
where program mutations capture defects that commonly arise during programming (a test suite's effectiveness is evaluated by measuring how many of these mutations are detected -killed- by the suite). Thus, such common faults would be repaired by simply reversing the mutations, leading to the motivation of our approach.  
While we use an extension of MuJava, \Stryker{} does not directly depend on it, and could be used with other mutation tools. Also, new mutation operators may be incorporated, extending the space of fix candidates, or existing mutators could be disabled.  

It is important to also mention that, as opposed to \cite{LeGoues+2012}, our technique performs a bounded search that is \emph{exhaustive}, and that does not depend on heuristic functions. 

\textbf{Fix Candidates Assessment.} 
\label{spec-based-candidate-assessment-description}
To assess the fix candidates produced by the technique, we employ two technologies. Given a fix candidate, we first check whether the candidate is promising by employing a run time analysis: we run the candidate on a number of (automatically) collected inputs, monitoring that the contract is not violated, using JML RAC \cite{Leavens+(2002)}. If the candidate does not pass this check, it can be straightforwardly discarded. On the other hand, if the candidate passes the run time checks, we submit it to the same (more expensive) analysis used for fault detection, i.e., SAT based bounded verification using TACO. This second analysis phase is then similar to that used in \cite{GMK11}.

Our approach is based on the use of contracts for the programs to be repaired, which may be thought as a limitation, compared to repair tools that use tests as acceptance criteria for fixes. However, for the class of bugs that we aim at repairing, tests tend to fall short as acceptance criteria, causing the acceptance of fixes that are in fact incorrect. More precisely, intra-statement mutators produce ``fine-grained'' mutations (compared to the candidates produced by tools like \emph{GenProg}), for which test suites, even strong ones, are weak in distinguishing spurious fix candidates from actual repairs. For instance, a faulty implementation of \texttt{smallest} in benchmark \emph{IntroClass} \cite{introclass-benchmark,LeGoues+2015}, is accompanied in the benchmark by a quite strong suite (it covers all branches in the program, and, when measured with \textsf{major} \cite{Just2014}, has a mutation score of 100\%); however, for our mutations of depth at most 2 (i.e., we perform up to two mutations in the program), 381 of the mutants produced from the faulty program pass all tests, but none is an actual fix.

\section{Program Repair with {\large \textsf{Stryker}}}
\label{stryker}

As described in the previous section, our approach to program repair essentially consists of mutating faulty code, by applying intra-statement program modifications, with the intention of producing a fix candidate that repairs the existing faults in the input (faulty) source code. However, the ability of this approach to repair bugs greatly depends on the mutation operators considered, which at the same time affects the efficiency of the repair process, especially taking into account that some bugs may require multiple mutations. 

As an example, let us consider method \texttt{getNode} shown in Alg.~\ref{sample}, whose purpose is to retrieve the $i$-th node in a singly linked list. Even in such small piece of code there are many possibilities for mutation-based defects (the defects our approach is, in principle, able to repair). Figure~\ref{faultsGetNode} lists some examples of such faults (obtained by mutating statements in \texttt{getNode}); mutations of statement in line $j$ are listed as $j_a, j_b, \ldots$. Notice that if the mutation operators considered for the repair process only modify arithmetic operators (e.g., replacement of an arithmetic operator by another, introduction or removal of short-cut increment and decrement operators, etc.), then method \texttt{getNode} with bug $10_a$ cannot be repaired. Also, depending on the available mutation operators, some bugs may only be corrected through multiple mutations. For instance, if mutation operators on reference-based expressions modify either the left-hand side of an assignment, or the right-hand side of an assignment, then fault $10_a$ requires at least two consecutive applications of mutation operators to be repaired.

\begin{figure}[t]
\small
$$
\begin{array}{rl}
5_a: & (\mathit{result} \mbox{ != }\mathsf{null}\ \&\&\ \mathit{current}\mbox{ != }\mathsf{null})\\
5_b: & (\mathit{result} == \mathsf{null}\ \&\&\ \mathit{current} == \mathsf{null})\\
5_c: & (\mathit{result} == \mathsf{null}\ ||\ \mathit{current}\mbox{ != }\mathsf{null})\\
6_a: & (\mathit{i} == \mathit{current\_index} + 1)\\
6_b: & (\mathit{i}\ \mbox{!=}\ \mathit{current\_index})\\
9_a: & \mathit{current\_index} = \mathit{current\_index} - 1\\
10_a: & \mathit{current.next} = \mathit{current}
\end{array}
$$
\normalsize
\caption{Some faults in method \texttt{getNode} that can be modeled as mutations.}\label{faultsGetNode}
\end{figure}


The number of mutation operators to consider and the number of lines to mutate affect how big the search space of fix candidates will be. Moreover, since we are considering code that may contain several faults, or faults that cannot be fixed with just a single mutation (e.g., faulty statements whose acceptable fixes require multiple mutations), it is often necessary to iterate the mutation process on the obtained mutants. This leads to a rapid explosion on the number of produced mutants. Table~\ref{mutExplosion} illustrates this explosion; it reports the number of mutants generated by mutating lines 5, 6, 9 and 10 in Alg.~\ref{sample} using 18 different mutators (some mutators produce multiple mutations). Notice how, as the number of mutation generations (i.e., number of times the mutation process is iterated) 
increases, the number of fix candidates grows quickly. 
This example shows that mechanisms that enable us to prune this search space are essential to make mutation-based program repair effective. Our approach is equipped with such pruning mechanism, which greatly contributes to its effectiveness, and is described in the next section. We now 
show how all the modules that constitute \Stryker{}, as depicted in Fig.~\ref{architecture}, are realized.

\paragraph{Fault Detection with {\large \TACO{}}}
\label{detection} 

To determine if a method under analysis is faulty we use the bounded verification tool \TACO{} \cite{Galeotti+2013}. \TACO{} receives as inputs a JML \cite{Burdy+2005} annotated method 
and bounds on the number of loop-unrolls and on the sizes of data domains. A translation maps JML-annotated code to an Alloy \cite{Jackson2006} model which, after being converted to a KodKod \cite{TJ07} model, yields a propositional formula that is solved using an off-the-shelf SAT-solver. The formula is obtained by conjoining:

\textbf{1.} requires($\mathrm{initial\_state}$), a propositional translation of the JML requires clauses, instantiated in the initial state.

\textbf{2.} $\mathit{Inv(\mathrm{initial\_state})}$, a propositional translation of the object invariant, instantiated in the initial state.

\textbf{3.} $\mathrm{code}(\mathrm{initial\_state}, \mathrm{final\_state})$, a propositional translation of the code under analysis.

\textbf{4.}  $!(\mathit{Inv}(\mathrm{final\_state}) \wedge \mathrm{ensures}(\mathrm{final\_state}))$, a propositional translation of the negation of the invariant and the ensures clauses, instantiated in the final state.

The outcome of the analysis using \TACO{} is an \emph{unsat} verdict (in case the code obeys its JML contract within the provided bounds), or an input exposing a contract violation.

\paragraph{Fault Localization}

For fault localization, we use an approach based on unsat cores. To retrieve unsat cores, unsatisfiable formulas are necessary. We build such formulas as follows. Let us suppose that a fault has been detected by \Stryker{}'s fault detection process. Then, the conjunction of formulas 1-4 in the previous paragraph has been found satisfiable, and a satisfying instance has been produced. This instance carries states $s_i$ and $s_f$ satisfying requires($s_i$), $\mathit{Inv(s_i)}$ and $!(\mathit{Inv}(s_f) \wedge \mathrm{ensures}(s_f))$, and a trace $t$ satisfying $\mathrm{code}(s_i, s_f)$) (including all intermediate program states). Then, the formula:
\begin{displaymath}
\textrm{requires}(s_i) \wedge \mathit{Inv(s_i)} \wedge t \wedge (\mathit{Inv}(s_f) \wedge \mathrm{ensures}(s_f))
\end{displaymath}
is unsatisfiable. The unsat core obtained for it highlights the cause of the inconsistency, i.e., the reason why the program failed on input $s_i$. Since this unsat core is input dependent, we collect various failing inputs and replicate this process on each of them. Then the statements of $t$ consistently highlighted by the unsat cores are identified as suspicious.

\paragraph{Mutants Generation}
\label{mutantsGeneration} 

In \Stryker{}, fix candidates are \emph{mutations}, generated using our extension of \muJava{} \cite{Ma+2005} (available in \cite{mujavapp-site}). We employ all method mutators available in \muJava{}, a total of 18 syntactic mutators.  
The most noticeable difference between our version of \muJava{} and the standard tool is the reimplementation of mutator PRV. This operation takes a navigation expression of the form $f_1.f_2.\dots.f_k$ and mutates it in all possible ways in which a method or field is added to the expression (in places where the typing allows for it), all possible ways of removing fields or methods from the expression that make the result type-consistent, and replacing a method/field in the expression by another that makes the resulting expression well typed. For instance, for expression \texttt{root.left.right}, mutants \texttt{root.right}, \texttt{root.right.right} and \texttt{root.left.left.right} would be produced, among many others.

As a result of a fault localization stage, a number of suspicious lines are identified. These statements are annotated with comments of the form \texttt{//mutGenLimit k}, where $k$ is greater than or equal to 0. These annotations mark the lines assumed to be buggy, and bound the number of times the mutation process can be applied to each statement. 

Since we aim at a bounded-exhaustive program repair approach, we must ensure that no feasible mutations are skipped during the search. Our implementation of \muJava{} provides a method \texttt{obtainMutants} that, given a method with \texttt{//mutGenLimit} annotations, returns for each annotated statement the applicable mutations. Notice that if we number the $k_i$ applicable mutations for statement $S_i$ using values 1 through $k_i$, and use 0 to denote that no mutation will be applied, we can describe each single (simultaneous) combination of applications of up to 1 mutation per mutable statement as an array of indices $[\ j_1\ |\ j_2\ |\ j_3\ |\ \cdots\ ]$ where $0 \leq j_i \leq k_i$. For example, an array $[\ 3\ |\ 0\ |\ 5\ |\ \cdots\ ]$ denotes a configuration where the first mutable statement will be mutated using its third mutation, the second mutable statement will not be mutated, and the third mutable statement will be mutated using its fifth mutation. We can then iterate through all these configurations by generating suitable arrays in lexicographical order. Therefore, we can easily traverse all mutants obtainable by mutating each (mutable) statement at most once. Since we are allowing to mutate single statements more than once, we need to consider such multiple mutations as well. A mutant obtained from a source method $M$, in which arbitrarily many mutations have been applied in each mutable statement, can be thought of as a mutation of a method $M'$, where $M'$ is obtained from $M$ by mutating each mutable statement at most once. We can represent the mutated statements from $M$ as an array $[\ m_1;m_2;m_3\ |\ p_1;p_2\ |\ q_1;q_2;q_3;q_4\ |\ \cdots\ ]$, where $m_i$, $p_j$ and $q_l$ are mutations to be applied on the source statements from $M$. Let $M'$ be the method obtained by applying mutations $m_1, p_1, q_1,\ldots$ on statements $S_1, S_2, S_3,\ldots$, respectively. The resulting method can then be obtained by applying mutations $[\ m_2;m_3\ |\ p_2\ |\ q_2;q_3;q_4\ |\ \cdots\ ]$ on $M'$. Notice that:
\begin{itemize}
\item All mutations, even those including the application of multiple mutators in a single statement, can be obtained in this way. Therefore, the method is complete (with respect to the faults class it targets).
\item Some mutations can be obtained in different ways. For instance, the effect of applying $m_1, p_1, q_1,\ldots$ and afterwards  $[\ m_2;m_3\ |\ p_2\ |\ q_2;q_3;q_4\ |\ \cdots\ ]$, is the same as applying $m_1, q_1,\ldots$ (notice that $p_1$ has been omitted) and afterwards applying $[\ m_2;m_3\ |\ p_1;p_2\ |\ q_2;q_3;q_4\ |\ \cdots\ ]$. In order to avoid considering repeated mutations we will hash generated mutants and remove those yielding collisions. 
\end{itemize}

\paragraph{Removing Wrong Fix-Candidates with JML-RAC} \label{RAC} 
RAC (Runtime Assertion Checker) \cite{Leavens+(2002)} is an application, part of the JML suite of tools, that allows one to jointly execute  a method and its JML contract (object invariants, requires and ensures clauses). We might remove spurious mutants resorting to \TACO{} (we will come back to this in the next paragraph), but \TACO{} analyses, while effective, are time consuming. Therefore, as a fast sieve to remove unsuitable mutants, we execute each mutant on the inputs obtained through \TACO{} in the fault detection or fix candidate assessment phases. 
Recall that these inputs exhibited failures in the source method. Any mutant that fails to comply with the original contract when executed on one of these inputs, cannot be a successful fix candidate and is immediately discarded as such. 



\paragraph{Assessing Fix Candidates with {\large \TACO{}}}
\label{assesWithTACO} 

If a mutant has successfully passed the execution of the collected inputs, we will use \TACO{} in order to determine if there are other inputs that violate the contract, and if that is not the case, we will consider this mutant a successful fix candidate. In case \TACO{} finds an input that makes the current mutant to violate the contract, two things happen: \emph{(i)} the mutant is discarded as a fix since it violates the contract, and \emph{(ii)} the input just found is added to the pool of inputs used for run time checking of other fix candidates. 

\section{Specification-Based Mutants Pruning}\label{pruningSection} 

As previously described, the amount of mutants (candidate program fixes) generated from a faulty program as part of the repair process  
grows in a geometric way. Then, the feasibility of the approach requires an effective mechanism to get rid of invalid mutants, i.e., to prune invalid candidates. Such pruning mechanism is what motivated the name of our tool, \Stryker, which comes from the mutants-obsessed character William Stryker, from X-Men. 

\medskip
``\emph{I've been working with mutants as long as you have, Xavier... but the most frustrating thing I've learned is that nobody really knows how many even exist or how to find them}."
\begin{flushright} 
William Stryker, X-Men 2.
\end{flushright} 

We now present our technique for pruning mutants that cannot lead to a fix. Notice that we will not only remove mutants that are not fixes (those are removed using run time checking and bounded verification, as explained in the previous section); we will remove mutants that, even if further mutations were applied, cannot lead to successful fixes.

\subsection{The Intuition Behind Mutants Pruning} 
\label{intuitionExplanation}
\begin{figure}[tb!]
\includegraphics[scale = 0.4]{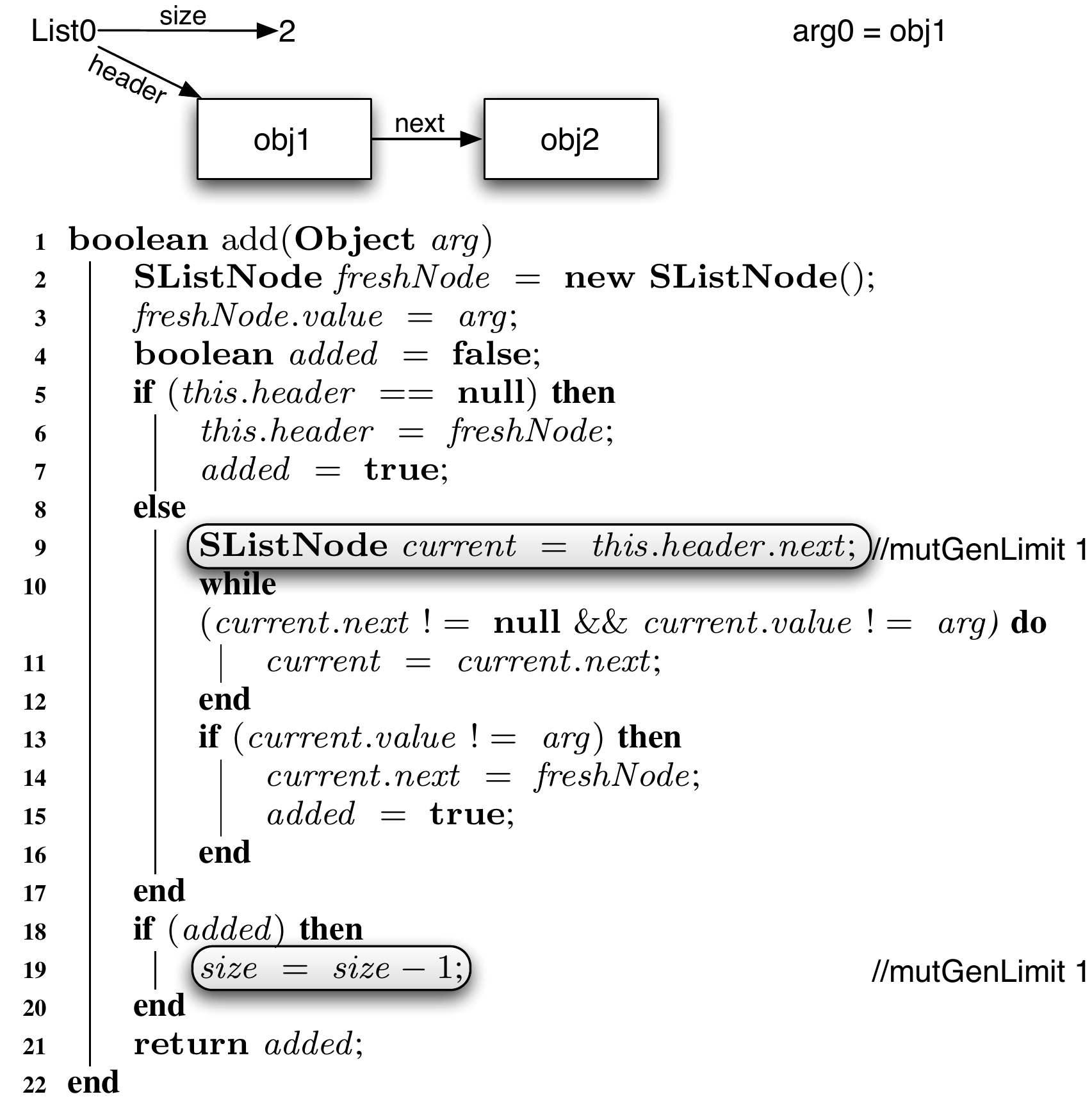}
\caption{Sample faulty code and an input found by \TACO\ exhibiting the faulty behavior.}\label{addFig}
\end{figure}

Let us convey the intuition behind our pruning technique by means of an example. Different faults in a program may relate to the computation of unrelated functionalities. For example, in a \texttt{Set} class that maintains a \emph{size} attribute and is implemented using a singly linked list, method \texttt{add} must perform the actual insertion, but also must establish the correct value for \emph{size}. Consider method \texttt{add} in Fig.~\ref{addFig}; in this method, two faults are present, one in line 9 (where the first list node is skipped) and the other in line 19 (the size is decremented instead of being incremented). An input state found by \TACO{}, which violates the method's contract, is given at the top of Fig.~\ref{addFig}.

Recalling our discussion in Section~\ref{mutantsGeneration}, for each mutation of statement 9 we have to iterate over all the mutations for statement 19. In particular, this is the case when statement 9 is not mutated or the applied mutation skips the first value from the input list; for all these cases, no matter which is the mutation considered for statement 19, the resulting mutant method will still violate the contract (notice that mutations related to \emph{size} will not fix the unrelated fault in statement 9).  The question is then

\medskip
\noindent
\emph{how can we automatically determine, without generating the mutations for statement 19, that the current mutation for statement 9 can be skipped?}
\medskip

We will begin by answering the question for this specific case, and in Section \ref{mutantsPruningTechnique} will generalize this solution to arbitrary methods. Let us consider the following JML-annotated version of the faulty \texttt{add} method\footnote{For the sake of clarity, we will omit considering at this point the object invariants.}:

{\small
\begin{verbatim}
    //@ requires this is List0;
    //@ requires arg is arg0;
    //@ ensures !add_ensures;
    boolean add(Object arg){
        method `add' code;
    }
\end{verbatim}
}

According to Section \ref{detection}, \TACO\ builds a propositional formula as the conjunction of (propositional translations of):
\begin{enumerate}
\item\label{firstItem} `this' is List0,
\item arg is arg0,
\item the (unrolled according to scope) code for method `add',
\item\label{lastItem} !!add\_ensures(final\_state) (which, after simplifying the double negation, becomes add\_ensures(final\_state)).
\end{enumerate}

Since the input state determined by \emph{List}0 and \emph{arg}0 led to a contract violation, add\_ensures(final\_state) must be false, and the propositional formula resulting of conjoining \ref{firstItem}--\ref{lastItem} is indeed unsatisfiable. Let us now focus on statement 19 from Fig.~\ref{addFig}: $\mathit{size} = \mathit{size} - 1$. If a mutation to (the right-hand side of) statement 19 is sufficient to fix the method, then, for the fault-exposing initial state, expression $\mathit{size} - 1$ is mutated into an expression $e$ whose value makes the post-condition hold. If such value would exist, we could refer to it by means of a fresh integer variable $i$, added as an input parameter to method \texttt{add}, and replacing statement 19 by $\mathit{size} = i$. Consider now the following JML-annotated method:

{\small
\begin{verbatim}
    //@ requires this is List0;
    //@ requires arg is arg0;
    //@ ensures !add_ensures;
    boolean add(Object arg, int i){
        statements 1-18;
        size = i;
        statements 20-22;
    }
\end{verbatim}

}

Analysis with \TACO\ searches for suitable values for \emph{this}, \emph{arg} and $i$ that will violate this contract. Notice that because of the negation in the ensures clause and the hardcoded values for \emph{this} and \emph{arg} in the requires clauses, this actually means finding suitable values of $i$ that make \texttt{add\_ensures} true.  \TACO\ will perform this search with the support of the SAT-solver, and return in this case an UNSAT verdict. This means that there is no such value for $i$ (and therefore no mutation for $\mathit{size} - 1$) that will make the code work starting from $\mathit{List}0$ and $\mathit{arg}0$. This allows us to conclude that it is safe to prune all the mutations on the right-hand side of statement 19. If, on the other hand, a SAT verdict would have been produced, that would mean that some value exists that allows the (modified) execution to complete correctly. Perhaps this value cannot be obtained via mutation; our conservative approach cannot distinguish this case and we will therefore iterate over the possible mutations.

Then, by adequately instrumenting the code and making a satisfiability check with \TACO{}, it is possible to determine whether the mutations of an expression can be skipped without compromising the fault correction procedure. 

\subsection{Mutants Pruning:  The Technique}\label{mutantsPruningTechnique}
We will begin this section by describing in Section \ref{instru} the instrumentation of the source code. In Section \ref{algorithm} we present the fault correction algorithm. Finally, in Section \ref{theos}, we will discuss the soundness of the approach.

\subsubsection{Instrumenting the JML-Annotated Source Code}\label{instru}
Let us assume that the JML specification of a method $M$ under analysis consists of an \texttt{invariant} clause (the object invariant for the class under analysis), a \texttt{requires} clause (precondition of the method under analysis) and an \texttt{ensures} clause (postcondition of the method under analysis). 
We will denote by $\mathit{pre}$ the conjunction of the object invariant and the requires clause. Similarly, we denote by $\mathit{post}$ the conjunction of the ensures clause and the invariant (the latter, instantiated on the final state).

As a result of the fault localization module, we assume some method statements include \texttt{//mutGenLimit k} annotations (with $k$ an integer greater than or equal to 0), bounding the number of \muJava\ mutations that can be applied to the statement. Let $S_1,\ldots,S_m$ be the statements in the source method whose mutGenLimit is greater than 0, listed in order, \emph{from the bottom} (i.e., $S_1$ is the last mutable statement and $S_m$ is the first mutable statement in the method).
The instrumentation then begins by unrolling loops as many times as prescribed by the scope. We will call the method obtained after loops are unrolled, $M_U$. Notice that statements that were mutable in $M$ and that occur inside a loop, now get repeated in each of the body replications resulting from the loop unrolling procedure. We will identify all these replications with their corresponding original statement. 

Assuming \TACO\ has been run on $M$ and a fault was detected, an input $I$ that makes the program fail, consisting of values \emph{args}0 for the method parameters and an object \emph{this}0 in case $M$ was not static (we assume this is the case; handling static methods is easier), has been produced. $M_U$ then receives the following contract (we will denote this JML-annotated version of $M_U$ by $M_U^{\mathit{JML, I}}$):

{\small
\begin{verbatim}
//@requires this == this0 && args == args0;
//@ensures !post;
\end{verbatim}
}

We now introduce the \emph{variabilization} technique. As explained in Section \ref{intuitionExplanation}, this introduces an abstraction from actual mutations that will allow us to prune infeasible mutants generations. We first define the different (depending on the statement kind) ways to variabilize, and afterwards present the notion of $k$-variabilization for a given method.

Given a mutable statement $S$ from $M$ ($S \in \{S_1,\ldots,S_m\}$) and a sequence $V$ of variables, we define the variabilization of $S$ as follows ($x$ denotes a variable, $f$ denotes a class field,  and $t, t_1, t_2$ denote terms):
\begin{itemize}
\item if $S$ is $x = t$, then for each replication $S_i$ of $S$ in $M_U$, $S_i$ is replaced by $x = \mathit{Var}_{S_i}$, where $\mathit{Var}_{S_i}$ is a fresh variable, and $V$ is extended with $\mathit{Var}_{S_i}$.
\item if $S$ is $t_1.f = t_2$, then for each replication $S_i$ of $S$ in $M_U$, $S_i$ is replaced by $\mathit{LHSVar}_{S_i}.f = \mathit{RHSVar}_{S_i}$, where $\mathit{LHSVar}_{S_i}$ and $\mathit{RHSVar}_{S_i}$ are fresh variables, and $V$ is extended with $\mathit{LHSVar}_{S_i}, \mathit{RHSVar}_{S_i}$.
\end{itemize}

Given a method $M$ and input $I$, its $k$-variabilization, denoted by $M_U^{\mathit{JML, I}}(k)$ is inductively defined on the value of $k$ ($k > 0$). Let us consider method $M$ and initialize the sequence of variables $V$ as the sequence of formal parameters of $M$. Method $M_U^{\mathit{JML, I}}(1)$ is defined as follows:
\begin{itemize}
\item its contract is the one from $M_U^{\mathit{JML, I}}$,
\item its body is obtained from the body of $M_U$ by variabilizing statement $S_1$ from $M$,
\item its arguments are the arguments from $M_U^{\mathit{JML, I}}$, plus the variables in the sequence $V$ obtained after the variabilization of statement $S_1$.
\end{itemize}

In order to define $M_U^{\mathit{JML, I}}(k)$ (for $k>1$) we assume we already calculated $M_U^{\mathit{JML, I}}(k-1)$. The resulting method is then characterized as follows: 
\begin{itemize}
\item its contract is the one from $M_U^{\mathit{JML, I}}(k-1)$,
\item its body is obtained from the body of $M_U^{\mathit{JML}}(k-1)$ by variabilizing statement $S_k$ from $M$,
\item its arguments are the sequence of variables in the sequence $V$ obtained from the ($k-1$)-variabilization of $M$, extended with the variables introduced along the variabilization of statement $S_k$.
\end{itemize}

\subsubsection{The Pruning Algorithm}\label{algorithm}

Algorithm $\mathtt{getFeedback}$ (Alg.~\ref{get-feedback}) computes the \emph{feedback} required by the pruning technique. This algorithm reports how many mutable statements can be skipped along the mutation process. Intuitively, if the $k$-variabilization of method $M$, when analyzed with \TACO, returns an unsat verdict, then no mutation of statements $S_1,\ldots,S_k$ may yield a fix candidate. Therefore, we can skip all such mutations. On the other hand, if the $k$-variabilization of method $M$ is satisfiable, this means that there might exist a mutation of statements $S_1,\ldots,S_k$ that fixes the bug (although perhaps the values assigned by the SAT-solver to the fresh variables cannot be denoted by program expressions, therefore skipping fewer unsuitable mutations than we might).

\begin{algorithm}[t]
\small
\textbf{Method} $\mathrm{getFeedback}$\;
\textbf{Input:} annotated method $M$\;
\textbf{Input:} $\mathit{args}$ for $M$ that expose a failure\;
\textbf{Output:} $k$ (number of mutable statements whose mutations can be skipped)
\Body{$k$ = 1\;
unsat? = true\;
\While{($k <= m$ \&\& unsat?)}{
$ M = M_U^{\mathit{JML, args}}(k)$\;
unsat? = \textsf{TACO}($M$)\;
$k = k+1$\;
}
\Return{k}\;
}
\normalsize
\caption{Computing the \emph{feedback} required along the pruning process.}\label{get-feedback}
\end{algorithm}

Algorithm \ref{alg-pruning} iterates over the mutations of a method $M$, skipping infeasible mutations with the aid of Alg.~\ref{get-feedback} (see Lines 26--32). Method $M$ is obtained from a queue of methods with pending mutations (Line 5). This queue is initialized with the original method under analysis, and its content evolves along the successive calls to Alg.~\ref{alg-pruning} (Lines 22,23) until the queue becomes empty or a successful fix is found. Recall that, as explained in Section \ref{mutantsGeneration}, mutations of a method $M$ with mutable statements $S_1,\ldots,S_m$ can be described via arrays of the form $[\ j_1\ |\ j_2\ |\ j_3\ |\ \cdots\ |\ j_m\ ]$, 
where $0 \leq j_i \leq k_i$ and $k_i$ ($1 \leq i \leq m$) is the number of mutations that can be applied to statement $S_i$.  A particular array $[\ j_1\ |\ j_2\ |\ j_3\ |\ \cdots\ |\ j_m\ ]$ describes the method obtained from $M$ by mutating statement $S_i$ ($1 \leq i \leq m$) using the $j_i$-th mutation from the sequence $\mathit{mut}_1,\ldots, \mathit{mut}_{k_i}$ of mutations applicable to $S_i$. We will iterate through these arrays in lexicographical order, with position 1 (the leftmost index in the array) being the least significative position.  Notice that array $[\ 0\ |\ 0\ |\ 0\ |\ \cdots\ |\ 0\ ]$ is the smallest one in the ordering, and  indicates that no mutation is applied. The next array in the ordering will then be $[\ 1\ |\ 0\ |\ 0\ |\ \cdots\ |\ 0\ ]$. Algorithm \ref{alg-pruning} will iterate over these arrays (Lines 6,7,28,29) following the ordering and skipping those arrays that method \texttt{getFeedback}  deems as unsuitable for becoming fix candidates. Most variables and methods invoked from Alg.~\ref{alg-pruning} have self-explanatory names. Method \texttt{advanceOneStartingAtIndex}($i$) (with $i \leq m$) adds 1 to the value stored in position $i$ of the array using carry in case the value previously stored was $k_i$ (the maximum possible value in position $i$). For example, for array $[\ 3\ |\ 2\ |\ 5\ |\ k_4\ |\ k_5\ |\ 2\ |\ 4\ ]$, \texttt{advanceOneStartingAtIndex}($3$) yields the array $[\ 3\ |\ 2\ |\ 5\ |\ 0\ |\ 0\ |\ 3\ |\ 4\ ]$.

\begin{algorithm}[t]
\small
\textbf{Method} $\mathrm{iterateWhilePruning}$\;
\textbf{Input:} $Q$, queue of methods with pending mutations to be applied\;
\textbf{Input:} set of inputs $S$ that satisfy $M$'s precondition\;
\textbf{Output:} $M'$ (first successful fix found), or ``not fixable" if no fix is possible.
\Body{
   Method $M$ = $Q$.pop()\;
   $k_1,\ldots,k_m$ = $M$.getNumberOfMutationsPerMutableStatement()\;
   ArrayIterator \emph{it} = new ArrayIterator($k_1,\ldots,k_m$)\;
   \textbf{boolean} candidateFound = false\;
   \While{(\textrm{it.\emph{hasNext()}} \&\& !\emph{candidateFound})}{
   	Method $M'$ = $M$.applyMutsAndUpdatePendingMuts(\emph{it}.next())\;
	\textbf{boolean} passesRAC? = false\;
	\For{(Input $i$ : $S$)}{
	   passesRAC? = passesRAC? $\|$ JML-RAC($M'$,$i$)\;
	}
	\If{(\emph{passesRAC?})}{
	   \textbf{boolean} passesTACO? = \TACO($M'$)\;
	   \If{(\emph{passesTACO?})}{
	      \Return{$M'$}\; 
	   }
	}
	\If{(!\emph{passesRAC?} $\|$ !\emph{passesTACO?})}{
	   \If{($M'$.\emph{furtherMutationsAreAllowed()})}{
	   	$Q$.add($M'$)\;
	   }
	   Input $I$ = failing input from JML-RAC or \TACO\;
	   \textbf{int} $k$ = \texttt{getFeedback}($M'$,$I$)\;
	   \eIf{($k < m$)}{
	   	\emph{it}.setToZeroAllPositionsInRange(0,$k-1$)\;
	   	\emph{it}.advanceOneStartingAtIndex($k$)\;
	   }{
	   \Return{``not fixable"}\;
	   }
	}
   }
}
\normalsize
\caption{Pruning along mutations traversal.}\label{alg-pruning}
\end{algorithm}

\subsubsection{Soundness}\label{theos} We will address soundness by answering two questions:
\begin{itemize}
\item Does the mutants enumeration procedure enumerate all mutants if pruning is disabled?
\item Does pruning only skip infeasible mutants?
\end{itemize}
Notice that a positive answer to these questions guarantees the soundness of the technique: all mutants are considered, and only unsuitable ones are pruned.

The first question can be positively answered by looking at Alg.~\ref{alg-pruning}, with Lines 25--32 removed. First, it is clear that given a method $M$ from $Q$, Alg.~\ref{alg-pruning} will iterate over all arrays, and therefore all mutants in which each mutable statement is mutated at most once, is correctly generated. If $M'$ (the current mutant), is not a fix candidate and still has pending mutations (Lines 21, 22), it is pushed in $Q$ for further mutation. This iterative process guarantees all the mutants will be produced. Termination is guaranteed because all the mutants that enter into $Q$ have at least one less pending mutation.

The second question can also be answered positively. A complete proof proceeds by induction on the number of variabilizations. We will focus on the base case, since the argument easily generalizes to the inductive step. Let us consider $M_U^{\mathit{JML, I}}(1)$. If the analysis with \TACO\ returns an unsat verdict, it means that there is no way of assigning values to statement $S_1$ such that will produce a valid path from input $I$. In particular, since mutations will adopt some of the values already considered by the underlying SAT-solver, no mutation of $S_1$ can fix the error exposed by input $I$.  It is therefore safe to skip such mutations.

\section{Evaluation}\label{eval}

Our evaluation consists of  
an experimental assessment of the effectiveness of \Stryker\ for fixing faults in a benchmark comprising several collection implementations \footnote{Instructions to reproduce the experiments available in \cite{stryker-site}.}. 
These classes, for which we have provided adequate JML contracts including \texttt{requires/ensures} clauses, loop variant functions and class invariants, are the following:
\begin{itemize}
    \item \textbf{SinglyLinkedList (SLList):} An implementation of singly linked lists. We consider methods \texttt{contains} for membership checking, \texttt{getNode} to retrieve the $i$-th element in the list, and \texttt{insert} to add a new object to the end of the list.

    \item \textbf{NodeCachingLinkedList (NCLL):} A caching, circular, double linked list implementation of interface \texttt{List} from the Apache package \texttt{commons.collections}. We consider methods \texttt{contains}, \texttt{insert} and \texttt{remove}. Method \texttt{remove} is particularly interesting due to the use of a cache for storing removed nodes, preventing unwanted garbage collection.

    \item \textbf{BinarySearchTree (BSTree):} A binary search tree implementation with methods \texttt{contains}, \texttt{insert} and \texttt{remove}.

    \item \textbf{BinomialHeap (BinHeap):} An implementations of priority queues using binomial heaps. We consider methods \texttt{findMin} (to retrieve the minimum element stored), \texttt{insert}, and \texttt{extractMin} to remove and return the least element.
\end{itemize}
\subsection{Experimental Setup}
Along the experiments we report in Section~\ref{actualData} we used a PC with Intel(R) Core(TM) i7-2600 CPU, running at 3.40Ghz and holding 8GB of RAM. We used GNU/Linux 3.2.0 as the OS. \Stryker{} was ran as an Eclipse Java project, using OpenJDK 1.7 as the underlying Java platform.

\subsection{Experimental Results}\label{actualData}

To measure the effectiveness of \Stryker{}, in exhaustive search mode (with no pruning) and with the pruning technique enabled, we proceed as follows. 
For each of the analyzed methods, we perform up to $4$ mutations, and randomly select 5 faulty programs for each number of mutations, between 1 and $4$ (i.e., we end up with 5 randomly chosen faulty programs with one bug, 5 faulty programs with 2 bugs, and so on, up to $4$ bugs). 
We mark each mutated line with the corresponding \texttt{//mutGenLimit}, accompanied with the exact number of mutations performed in the line. That is, we evaluate \Stryker{} with and without pruning under the assumption of ``perfect'' fault localization information (see later on in this section for repair under imprecise fault location information). For each experiment, we set the timeout to 10 hours.
Since, for every number $i$ of bugs between 1 and $4$ we have various randomly chosen faulty versions of each method, we report the minimum, maximum and average time consumed to find a fix (resp. number of fix candidates visited) for the $i$ bugs, comparing \Stryker{} in its ``no-pruning'' and ``pruning'' configurations. 
These charts show that, as the number of bugs in the code is increased, in general the pruning tends to provide increasingly significant benefits. It is important to observe that the running times, although better (and significantly better in a good number of cases), do not improve at the same rate as the state space reduction, in these case studies. This has to do, in principle, with the overhead caused by the additional SAT queries required for the pruning approach. As a concrete example,  consider method \texttt{BinomialHeap.insert} (80 loc). Notice how, for one bug, the number of visited structures and running times are essentially the same (in general, for one bug, the pruning overhead is not notorious, with the exception of the case \texttt{BinomialHeap.find}). As the number of bugs is increased, pruning greatly reduces the search space. In particular, for 4 bugs, \Stryker{} with pruning visited 769 structures in the \emph{worst case}, while the tool without pruning visited 4642 in the \emph{best case}; running times, while did not improve at the same rate, in average provided important time savings (655146 milliseconds in average with pruning, against 3629759 in average without pruning).



\begin{figure*}[h!]
    \includegraphics[width=18cm]{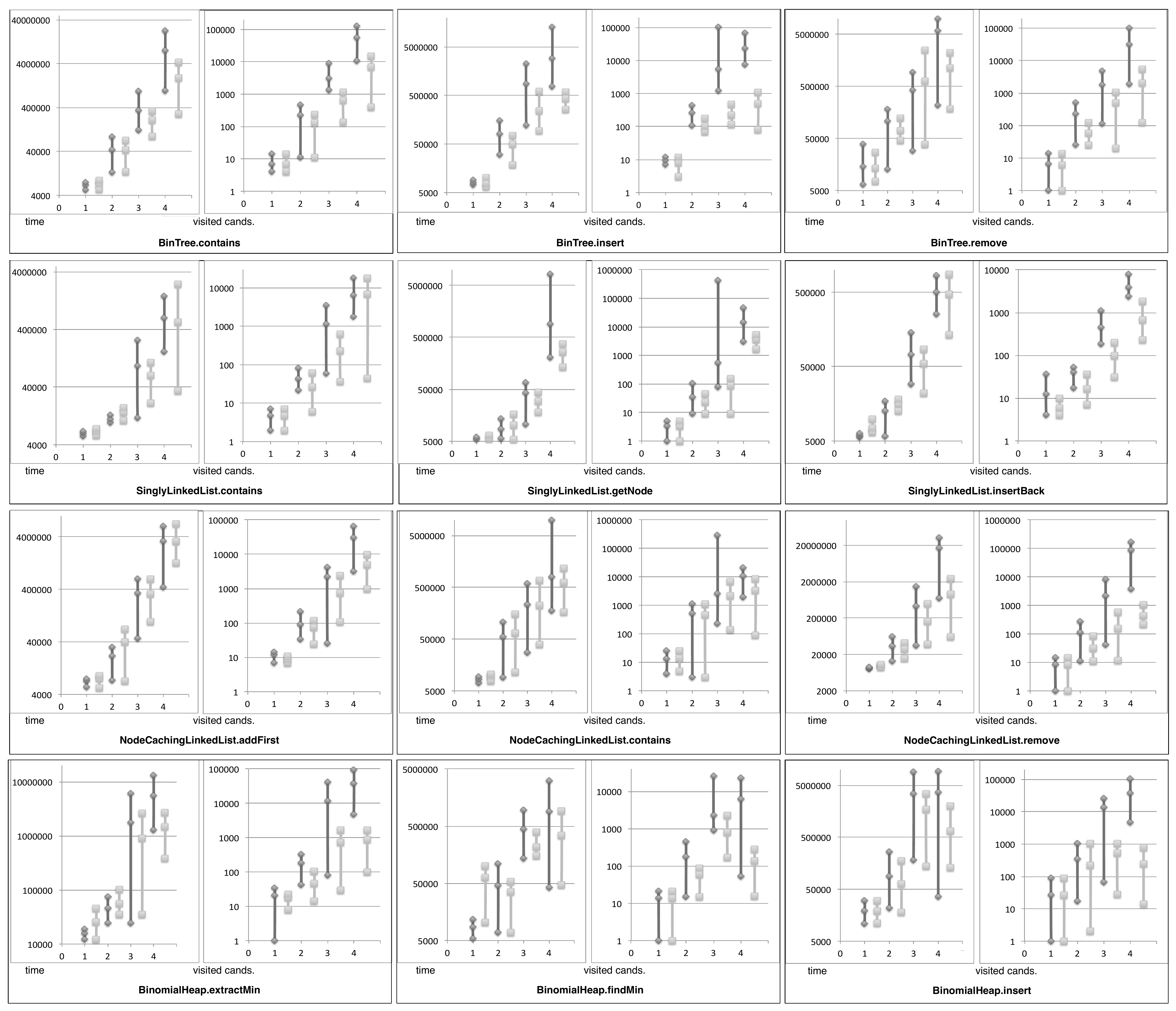}
    \caption{Experimental evaluation of \Stryker{}, both in time for repair and candidates visited, as number of bugs in code increases. Each line reports minimum, maximum and average. Dark lines correspond to exhaustive repair; light lines correspond to pruning strategy.}
    \label{resultCharts}
\end{figure*}

\Stryker{}'s architecture makes it essentially agnostic to the fault localization approach employed prior to repair (as opposed to other tools, e.g., \emph{GenProg}, for which fault localization is an active part of the fault repair process). The previous experiments, run under the assumption of perfect fault localization, allowed us to evaluate \Stryker{} independently of any specific fault localization. However, since fault localization mechanisms are known to be imprecise (and the one based on unsat cores used in this paper is no exception), evaluating \Stryker{}'s performance in such imprecise fault location information scenarios is definitely relevant. We ran a number of additional experiments with our benchmark, in which the unsat core based fault localization module either identified a superset of the actual faulty program locations, or missed at least one faulty program location (making the corresponding program, in the latter case, not repairable by the technique). As it may be expected, in cases in which fault localization identified a superset of the actual faulty locations, \Stryker{}'s space of candidates grows accordingly, and its efficiency is diminished proportionally to how the fix candidates space grew. On the other hand, in cases in which fault localization missed faulty lines, \Stryker{}'s pruning mechanism generally behaved efficiently, largely outperforming \Stryker{} without pruning. For instance, for a faulty version of \texttt{remove} from class \texttt{NodeCachingLinkedList} with $4$ bugs in which fault localization missed one of the faulty lines, it took \Stryker{} with pruning 59.5 seconds to exhaust the state space (visiting 61 candidates), while the tool without pruning spent 185.4 seconds to exhaust the state space (visiting 960 candidates). 


\section{Related Work and Discussion}\label{related}

Fault correction has become over the last few years a very active research topic. 
We will begin by comparing with closely related work. Debroy and Wong \cite{DW10} use mutation to compute fixes using a similar motivation to ours. They use \emph{first-order} mutants (only one mutation is applied), as a means to circumvent mutants explosion. Therefore, their technique misses the power offered by combining mutations, that our tool exploits. Gopinath et al. \cite{GMK11} do specification based code repair (as we do), 
but they only consider a very limited set of fix candidates (expressions of the form $v.f_1.\cdots.f_n$), allowing their tool to partially circumvent the mutants explosion problem.  They solve existentially quantified formulas representing program expressions to guide the search for fixes using \emph{satisfiable} cases, as opposed to our program variabilizations, that enable us to prune portions of the space of fix candidates in \emph{unsatisfiable} cases. 

Weimer et al.~\cite{WNLF09} present GP, a tool for patch synthesis through genetic programming. The authors recognize mutants explosion as an impediment:

\medskip
\emph{``the number of possible changes is still huge, and this has been a significant impediment for GP in the past.''} \cite{WNLF09}
\medskip

They limit mutants generation by ``\emph{adopting existing code from another location in the program}''. While this may be a correct decision in many cases, it makes the technique incomplete. Unlike \cite{WNLF09}, \Stryker{} will only discard mutants that cannot lead to a program fix. On the other hand, GP uses expressive ``patches" that can fix faults that \Stryker{} will not fix due to the adopted notion of mutation. Kim et al.~\cite{KNSK13} propose a modified version of \cite{WNLF09} where 10 templates are used to fix faults, therefore discarding many other variations on the code under analysis. A limitation of this technique, that is partially shared by \cite{WNLF09}, is that some patches rather than fixing faults, seem to \emph{mask} said faults. This is the case for instance with patterns \texttt{Null Pointer Checker}, \texttt{Range Checker} and \texttt{Class Cast Checker}. For a thorough discussion of this work see \cite{M14}. Martinez and Monperrus \cite{MM13} carefully analyze the shape of changes done during fault correction, and identify 20 meaningful change categories. \Stryker{} can be applied for the correction of those faults that can be fixed by changes that do not require adding or removing statements to the program. 

\Stryker\ falls within the category of specification-based fault correction tools, and uses bounded exhaustive analysis to assess fix candidates, while many other approaches (e.g., \cite{WNLF09}) use testing as the underlying fix candidate assessment technique. The exhaustive analysis provided by \Stryker\ is more conclusive for detecting faults and discarding spurious fix candidates than testing. Indeed, various tools for program repair that employ testing as acceptance criteria for program fixes have been shown to produce spurious (incorrect) repairs \cite{Qi+2015}. 
Furthermore, while testing is more scalable than bounded verification, and may be effective as an acceptance criterion for fix candidates in some program repair approaches, in our context of bounded exhaustive exploration of candidates obtained from ``finer-grained'' intra-statement mutations, testing is intrinsically weak as acceptance (see Section~\ref{spec-based-candidate-assessment-description} for an example). 
Specification-based tools have a stronger and more precise acceptance criterion, that is essential to our process, for the class of faults we aim at repairing.  
Even for coarse grained mutations, testing as acceptance criterion for fixes shows limitations (see \cite{Qi+2015}). Also, in \cite{autofix-site}, a few fixes obtained by PAR and GenProg are reported, some of which are actually disabling functionality, in accordance with the results in \cite{Qi+2015}. 




%



The analysis performed by \TACO{}, the verification tool underlying \Stryker{}, depends on bounds for data domains and loop unrolls. Therefore, faults might go undetected in the initial phase, or fix candidates might be considered successful fixes even though they contain faults that cannot be detected within the used analysis bounds. Another limitation is the fact that the mutant generation process easily generates a large amount of mutants. 

A particularly successful approach to fault correction is that of GenProg. While GenProg is able to deal with large programs, the technique cannot deal well with several bugs. Basically, the presence of several bugs in the code affects the fitness of candidates, making the candidates population maintained by GenProg's evolutionary computation approach unstable. 


\section{Conclusions and Further Work}\label{conclu}

We presented \Stryker{}, a tool implementing a novel technique for program repair, that considers an ample set of intra-statement syntactic operations, and explores fix candidates exhaustively up to a provided bound. This technique combines run time analysis and SAT-based bounded verification for fix candidate assessment, and is able to repair contract-equipped faulty programs with several bugs. Moreover, we introduced as part of the tool a pruning technique that is able to discard large sets of infeasible fix candidates, by checking whether particular ``partial'' fix candidates (that mutate some suspicious statements) can lead to fixes or not, through the use of additional SAT queries. We evaluated our tool on a number of Java collection classes implementations, adequately equipped with corresponding contracts, showing the effectiveness of the technique.

The introduced pruning technique enables us to reduce the state space of candidates significantly, leading to an increased scalability. 
Our approach is heavily based on SAT solving. Since SAT solving is a very active and competitive field, and improvements in this area are constantly produced, our repair technique is likely to show better running time profits as SAT solvers become more efficient. 

\Stryker{}'s architecture and program repair approach makes it less coupled to particular tools and mechanisms to produce fix candidates (i.e., the supported mutation operations), and to localize faults. This enables, on one hand, the possibility of incorporating new (or alternative) mutation operators, changing the class of faults that the tool is able to repair. For instance, one may define mutations that introduce new statements, thus broadening the class of repairable faults. Also, the technique does not depend on any particular fault localization tool or approach (we used a prototypical fault localization technique that exploits unsat cores), leaving a choice for alternative localization strategies, e.g., \cite{JoseMajumdar2011,Moon+2014}, that could even be selected depending on the domain of the program to be repaired. 

There are various lines that we plan to explore as future work. \Stryker{}'s approach to fixing, presented as a search problem, leaves open the possibility of using different alternative strategies for performing this search, including heuristic search ones. For instance, while currently there is no particular order in treating different mutations to a same line of code, one may define reasonable priorities between mutators, to accelerate the time to find successful fixes. Also, although we adopted a bounded exhaustive search for program fixes, one may also resort to non-exhaustive explorations of the fix candidates space, e.g., via a genetic algorithm. 


\end{document}